\documentclass[acmsmall]{acmart}

\usepackage{booktabs} 
\usepackage{wrapfig}
\usepackage{enumitem}
\usepackage{gensymb}

\usepackage[ruled]{algorithm2e} 

\SetAlFnt{\small}
\SetAlCapFnt{\small}
\SetAlCapNameFnt{\small}
\SetAlCapHSkip{0pt}

\usepackage{xspace}
\makeatletter
\DeclareRobustCommand\onedot{\futurelet\@let@token\@onedot}
\def\@onedot{\ifx\@let@token.\else.\null\fi\xspace}
\def\eg{\emph{e.g}\onedot} 
\def\ie{\emph{i.e}\onedot}

\def\wrt{w.r.t\onedot} 
\def\etal{\emph{et al}\onedot}
\makeatother

\usepackage[makeroom]{cancel}

\newcommand{\dt}{{\Delta t}}

\newcommand{\btheta}{\boldsymbol{\theta}}

\newcommand{\bX}{\mathbf{X}}

\newcommand{\bp}{\mathbf{p}}

\newcommand{\bv}{\mathbf{v}}
\newcommand{\bx}{\mathbf{x}}

\newcommand{\bq}{\mathbf{q}}

\newcommand{\bz}{\mathbf{z}}

\newif\ifcomment
\commenttrue

\begin{document}

\title{Self-supervised Learning of Latent Space Dynamics}

\author{Yue Li}
\affiliation{%
  \institution{ETH Z{\"u}rich}
  \country{Switzerland}
}
\email{yue.li@inf.ethz.ch}

\author{Gene Wei-Chin Lin}
\affiliation{%
  \institution{Meta Reality Labs}
  \country{Canada}
}
\email{genelin@meta.com}

\author{Egor Larionov}
\affiliation{%
  \institution{Meta Reality Labs}
  \country{USA}
}
\email{elrnv@meta.com}

\author{Alja\v{z} Bo\v{z}i\'{c}}
\affiliation{%
  \institution{Meta Reality Labs}
  \country{Switzerland}
}
\email{aljaz@meta.com}

\author{Doug Roble}
\affiliation{%
  \institution{Meta Reality Labs}
  \country{USA}
}
\email{droble@meta.com}

\author{Ladislav Kavan}
\affiliation{%
  \institution{Meta Reality Labs}
  \country{Switzerland}
}
\email{
lkavan@meta.com}

\author{Stelian Coros}
\affiliation{%
  \institution{ETH Z{\"u}rich}
  \country{Switzerland}
}
\email{stelian.coros@inf.ethz.ch}

\author{Bernhard Thomaszewski}
\affiliation{%
  \institution{ETH Z{\"u}rich}
  \country{Switzerland}
}
\email{bthomasz@ethz.ch}

\author{Tuur Stuyck}
\affiliation{%
  \institution{Meta Reality Labs}
  \country{USA}
}
\email{tuur@meta.com}

\author{Hsiao-yu Chen}
\affiliation{%
  \institution{Meta Reality Labs}
  \country{USA}
}
\email{hsiaoyu@meta.com}

\renewcommand\shortauthors{Li, Y. \etal}

\begin{abstract}
Modeling the dynamic behavior of deformable objects is crucial for creating realistic digital worlds.
While conventional simulations produce high-quality motions, their computational costs are often prohibitive. 
Subspace simulation techniques address this challenge by restricting deformations to a lower-dimensional space, improving performance while maintaining visually compelling results. However, even subspace methods struggle to meet the stringent performance demands of portable devices such as virtual reality headsets and mobile platforms.
To overcome this limitation, we introduce a novel subspace simulation framework powered by a neural latent-space integrator. Our approach leverages self-supervised learning to enhance inference stability and generalization.
By operating entirely within latent space, our method eliminates the need for full-space computations, resulting in a highly efficient method well-suited for deployment on portable devices. We demonstrate the effectiveness of our approach on challenging examples involving rods, shells, and solids, showcasing its versatility and potential for widespread adoption.
\end{abstract}

%

\setcctype{by}
\acmJournal{PACMCGIT}
\acmYear{2025} \acmVolume{8} \acmNumber{4} \acmArticle{57} \acmMonth{8} \acmPrice{}\acmDOI{10.1145/3747854}


\maketitle
\section{Introduction}
\begin{figure}
    \centering
    \includegraphics[width=0.9\linewidth]{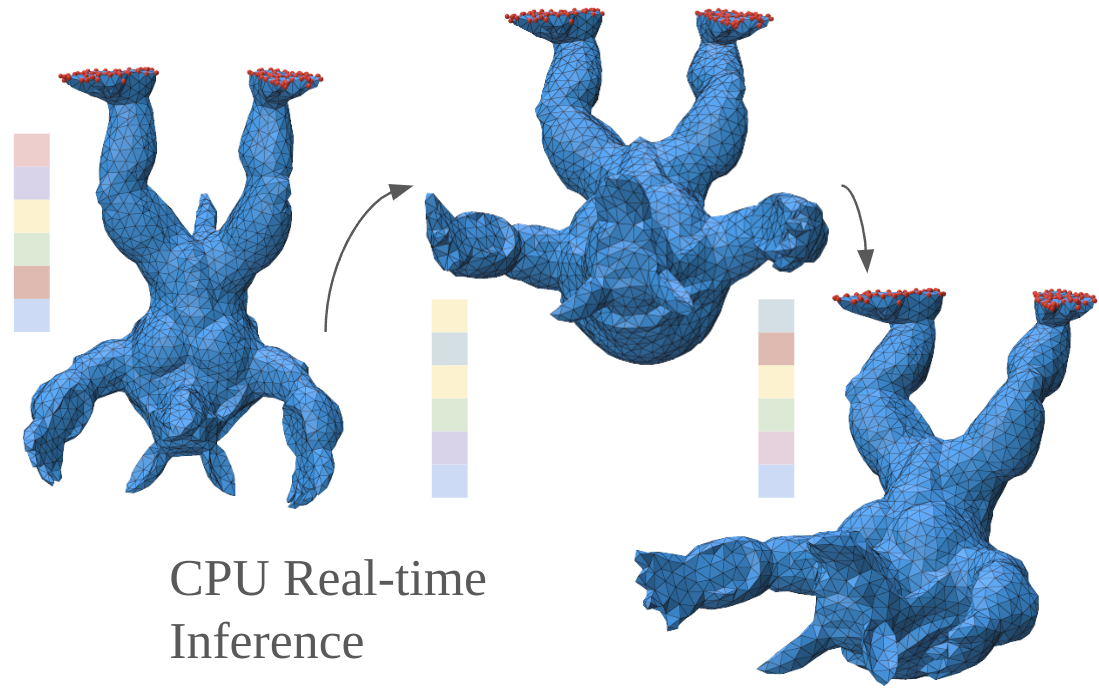}
    \caption{\textbf{Real-time Simulation:} We propose a novel latent space integrator that allows for robust and stable CPU-real-time (0.489 ms per frame) autoregressive inference of dynamic motions. Boundary conditions are shown as red spheres and the colored squares indicate low-dimensional latents.}
    \label{fig:teaser}
\end{figure}

Modeling the dynamic behavior of elastic objects is essential for creating realistic virtual environments. Physics-based simulations can produce highly compelling results, capturing the complex dynamics induced by interactions between digital humans and virtual objects. However, these simulations are computationally expensive, and even the most advanced methods struggle to meet the stringent real-time requirements of computer games, telepresence, or virtual reality applications.

Subspace simulation methods offer a promising alternative, providing an efficient trade-off between speed and generality. By focusing on the most likely object deformations, subspace approaches can accelerate simulations by orders of magnitude. However, most methods still require frequent evaluations of full-space energies and their derivatives during time stepping. These full-space evaluations require significant computational resources, precluding their deployment on compute-limited devices. As a result, achieving true real-time performance remains elusive even with subspace simulation methods.
In this work, we present a novel approach for neural simulation of deformable objects with natural dynamics. Unlike previous work, our method operates entirely in latent space, removing the need to evaluate full-space energies and their derivatives. Our two-step process begins with training an autoencoder on data sequences, followed by training a multilayer perceptron to predict an implicit integration step using initial positions and velocities in latent space. We build our method on a self-supervised learning approach that uses the variational implicit Euler potential as a loss function. We demonstrate that this self-supervised approach significantly improves generalization when compared to conventional supervised methods.
To ensure stable autoregressive inference, we propose a novel data augmentation strategy that adds noise to single-step predictions during training. This approach leads to improved stability for long roll-outs while avoiding multi-step predictions during training. We further identify a data imbalance issue that arises from physics-based self-supervised learning and propose a simple yet effective normalization strategy.
\par
We evaluate our method on a diverse set of examples that involve dynamic motions of elastic rods, shells, and solids. The results demonstrate that our latent-space integrator produces stable and plausible animations at a fraction of the computational cost required by full-space simulations.

In summary, our contributions are 
\begin{itemize}
    \item a novel method for implicit integration of elastodynamics equations in latent space using self-supervised learning.
    \item a subspace simulation method that operates exclusively in latent space during inference.
    \item a dedicated data augmentation and balancing method for improved stability and generalization.
\end{itemize}

\section{Related Work}

\paragraph{Subspace Simulation in Graphics}
Subspace simulation techniques have been extensively explored in the field of computer graphics to efficiently model complex dynamic systems~\cite{sharp2023data,Eftychios12FEM}, e.g. shells~\cite{zheng2024proxy}, cloths~\cite{hahn2014subspace,li2023subspace, holden2019subspace, de2010stable}, deformable bodies~\cite{barbivc2005real,harmon2013subspace,wang2015linear,wu2015unified,chen2023model,zheng2024multi}, animated characters~\cite{barbivc2009deformable,barbivc2012interactive,jacobson2012fast} and  fluids~\cite{treuille2006model,kim2013subspace}.
A common strategy is to construct subspace bases from the elastic energy Hessian at the rest state.
Early works focus on linear modal analysis~\cite{pentland1989good,hauser2003interactive,sifakis2012fem} several extension to the nonlinear setting have recently been proposed \cite{wang2024neural,benchekroun2023fast,duenser2022nonlinear}.

%
\par
While the eigenvectors of the elastic energy Hessian might be informative enough to construct reduced bases for deformable solids, hair exhibits a much less structured behavior. Arguably the most widely used  approach for hair simulation is based on \textit{guided strands}~\cite{chai2014reduced,lyu2020real,hsu2024real}, which involves simulating a select group of guide hairs from which the remaining hairs are interpolated. Hair meshes~\cite{yuksel2009hair,wu2016real} are an alternative approach which also allows for real-time rendering~\cite{bhokare2024real}. However, this approach requires (often manually) precomputed hair meshes with fixed topologies. Furthermore, this method is not appropriate for low-end devices as it requires dedicated GPU resources with mesh shader functionality, which is often absent from mobile hardware. 
To accommodate different types of deformable objects, 
we construct the map from reduced space to full space using deep neural networks, namely autoencoders.
%

\paragraph{Physics-informed Learning}
There has been an ongoing trend of leveraging physics-based loss functions to ground data-driven methods in computer vision and graphics~\cite{li2021deep,bertiche2022neural,santesteban2022snug,bertiche2020pbns,modi2024simplicits,chentanez2020cloth}. Early work performs physics-based simulation on the fly to either generate physics-based targets as supervisions~\cite{li2021deep} or correct self-intersections during reconstruction~\cite{halimi2022pattern}. Recent endeavors aim to learn the time integration function directly. In the supervised setting, Pfaff \etal~\shortcite{pfafflearning} and Sanchez-Gonzalez \etal~\shortcite{sanchez2020learning} employ graph neural networks to simulate particles and meshes. Similar to our work that adopts a self-supervised learning strategy, Bertiche \etal~\shortcite{bertiche2020pbns,bertiche2022neural}, Santesteban \etal~\shortcite{santesteban2022snug}, and Grigorev \etal~\shortcite{grigorev2023hood,grigorev2024contourcraft} learn quasi-static and dynamic integrators for cloth driven by articulated characters. Most recently, Stuyck \etal~\shortcite{stuyck2025quaffure} propose a self-supervised strategy for learning hair quasi-statics.
Whereas previous work focuses on learning full-space integrators, our method is the first to apply the self-supervised training paradigm in learning subspace integrators. We demonstrate that the resulting network is extremely lightweight and leads to real-time CPU performance.

\paragraph{Reduced Modeling with Neural Networks} There has been a recent surge in using deep neural networks as smooth and compact representations for reduced modeling. Zong~\etal~\shortcite{zong2023neural} leverage neural networks as implicit reduced representations for stress fields and Chen~\etal~\shortcite{chen2023model} model the deformation map for material point methods using neural representations. Utilizing deep neural networks directly on position levels for model reduction but working with continuous fields, Chen~\etal~\shortcite{chen2022crom} and Chang~\etal~\shortcite{chang2023licrom} learn smooth deformation maps from undeformed fields to deformed ones. Pan~\etal~\shortcite{pan2023neural} further leverage implicit representations for spatial-temporal data. In a similar vein but closely related to our work, Lee and Carlberg~\shortcite{lee2020model}, Fulton~\etal~\shortcite{fulton2019latent}, Zhang~\etal~\shortcite{zhang21dynamic} and Shen~\etal~\shortcite{shen2021high} use autoencoders to condense high dimensional position vectors into a low dimensional latent space. Additionally one could further improve the convergence by enforcing Lipschitz continuity~\cite{lyu2024accelerate}.
Similar in that we also resort to the autoencoder to construct a reduced subspace, we demonstrate that a physics-based integrator can be learned directly in this latent space.

\section{Method}
\begin{figure*}[h]
    \centering
    \includegraphics[width=\linewidth]{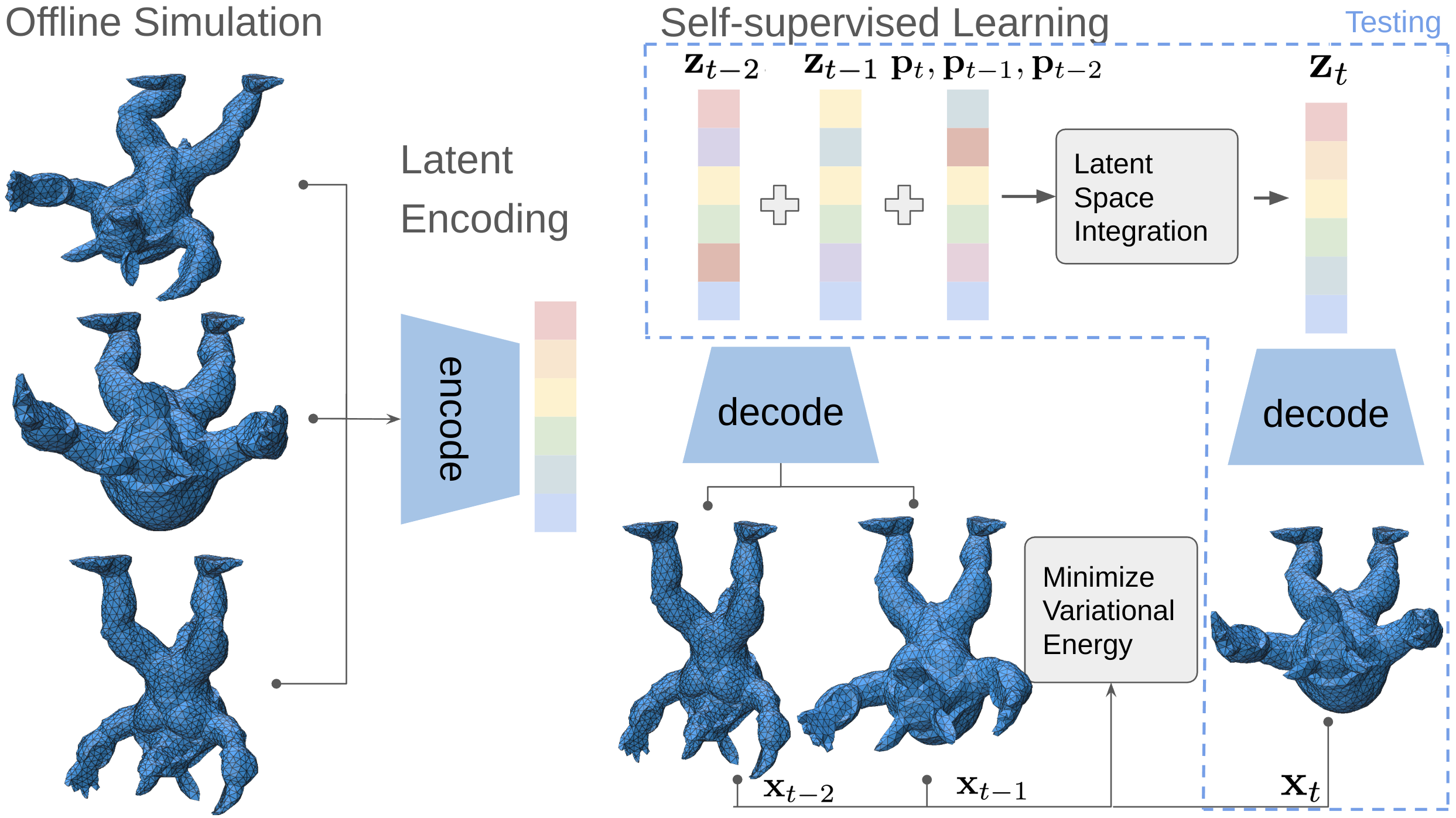}
    \caption{\textbf{Method Overview:} Our approach encodes full-space simulation data into low-dimensional latent vectors. We then learn an integrator in latent space in a self-supervised setting by minimizing the incremental potential in full space. During testing, as indicated by the blue dashed region, our integrator does not require any full-space computation and therefore runs in real-time on CPU.}
    \label{fig:overview}
\end{figure*}

Our approach learns a time integration scheme in a nonlinear latent space (see Fig.~\ref{fig:overview}). 
To construct the latent space, we begin by generating full-space motion data using an off-the-shelf simulation method (Sec.~\ref{Sec:full_space_simulation}). 
We then construct a low-dimensional nonlinear subspace from these data using autoencoders (Sec~\ref{Sec:latent_space_reconstruction}).
At the core of our approach, we learn an integrator in the subspace in a self-supervised manner by minimizing the full-space incremental potential (Sec~\ref{Sec:latent_space_learning}). 
\subsection{Full Space Simulation}
\label{Sec:full_space_simulation}
We initiate our approach by defining the full space energy that describes the physical system. This energy serves two key purposes: it generates the training data and supervises the training of the subspace integrator.
\par
To enable robust simulation, we employ implicit Euler as our time-stepping strategy. Following~\cite{martin2011example}, we cast the nonlinear root-finding problem into an optimization problem. At each simulation time step, we find the end-of-time-step position $\bx_t$ by minimizing the objective function
\begin{equation}
\begin{split}
    E_{\mathrm{total}}  &= E_{\mathrm{inertial}}(\bx_t, \bx_{t-1}, \bx_{t-2}) \\
    &+ E_{\mathrm{elastic}}(\bx_t, \bX) + E_{\mathrm{external}}(\bx_t) \ ,
\end{split}
\end{equation}
where $\bX$ denotes the rest state positions. These energies follow standard practices in graphics, $E_{\mathrm{inertial}}$ computes the inertial energy given three consecutive positions, $E_{\mathrm{external}}$ computes the energy due to external forces. The elastic potential $E_{\mathrm{elastic}}$ differs across applications. For rods, we use the stretching and bending models from \citet{bergou2008discrete}. We use discrete elastic shells~\cite{grinspun2003discrete} to model the material behavior of cloth, and linear tetrahedral elements with the StVK energy~\cite{sifakis2012fem} for solids. 

%
\par
We minimize this objective energy using Newton's method with backtracking line search~\cite{nocedal1999numerical}. To ensure robust convergence, we project local Hessian blocks to be positive-definite~\cite{teran2005robust}. Boundary conditions are imposed by removing the corresponding degrees of freedom from the system and setting the displacements to prescribed values. 
%
 
\subsection{Latent Space Representation}
\label{Sec:latent_space_reconstruction}
We utilize autoencoder as a convenient tool to compress full-space simulation data to an efficient nonlinear low-dimensional subspace.
We learn this nonlinear map by minimizing the reconstruction loss in full space

\begin{equation}
    \min_{\btheta_{d}, \btheta_e} \ \sum_t || \mathcal{D}(\mathcal{E}(\bx_t ; \btheta_{e}) ; \btheta_{d}) - \bx_t ||^2_2 \ ,
\end{equation}
where  $\btheta_{e}$ and $\btheta_{d}$ are the network parameters for the encoder $\mathcal{E}$ and decoder $\mathcal{D}$. We further refer to the output of the bottleneck layer $\bz_t := \mathcal{E}(\bx_t ; \btheta_{e})$ as latents. To remove high-frequency artifacts in the reconstruction, we initialize the first layer of the encoder and the last layer of the decoder with PCA bases following~\citet{fulton2019latent}. 
We encode vertex positions relative to Dirichlet boundary conditions. For rods, we leverage root relative encodings, \ie, every vertex along a given rod is coded by the relative distances to the root. We exclude root vertices from the latents and simply apply the boundary condition values whenever decoded to full states. For our cloth and solid examples, we encode the mesh vertices relative to the mean position of the Dirichlet vertices.

\subsection{Latent Space Self-supervised Learning}
\label{Sec:latent_space_learning}
To achieve real-time performance, we learn an integrator $\mathcal{I}$ that does not require any evaluation in full space during inference.
Similar to a first-order numerical integrator in full space, our integrator also takes the previous two states (latents) and the boundary condition history as inputs and predicts the corresponding latent codes for the current frame, 
\begin{equation}
    \bz_t = \mathcal{I}(\bz_{t-1}, \bz_{t-2}, \bp_t, \bp_{t-1},\bp_{t-2}; \btheta_{\mathcal{I}}) \ ,
\end{equation}
where $\btheta_{\mathcal{I}}$ denotes the weights and biases of the integrator network and the parameters $\bp_{t}, \bp_{t-1}, \bp_{t-2}$ are boundary conditions for three consecutive steps. For instance, for the rod swing example in Fig.~\ref{fig:rod_translation}, these are the three-time-step velocities of the moving cube.
Our integrator predicts new frames directly with a time step $\Delta t = 1/30$ seconds per frame.
\par
Given that we have ground truth latents available, one alternative is to adopt a supervised learning strategy to penalize the mismatch between the network prediction and the ground truth data in the $L_2$ sense. This approach, albeit simple, is inclined to overfitting and often fails to generalize~\cite{ma2023learning,zehnder2021ntopo}; see also Fig.~\ref{fig:comparison_supervised}.
\par
Inspired by recent advances in physics-informed learning, we propose a novel self-supervised learning strategy. While there isn't a closed-form functional defined in latent space, we can decode the latents to their full space configurations and then minimize the variational implicit Euler loss. Therefore, the loss function we used for learning the integrator is identical to the objective function that we minimize during full space simulation, except that now the optimization variables are neural network weights and biases, $\btheta_{\mathcal{I}}$, instead of positions. Denoting the inputs of the latent integrator as $\bq$, our loss function reads
\begin{equation}
\label{eqn:training}
\begin{split}
    \mathcal{L} (\btheta_I)  &= E_{\mathrm{inertial}}(\bx_t(\bz_t(\bq;\btheta_{\mathcal{I}})), \bx_{t-1}, \bx_{t-2})\\ &+ E_{\mathrm{external}}(\bx_t(\bz_t(\bq;\btheta_{\mathcal{I}})) +E_{\mathrm{bc}}(\bx_t(\bz_t(\bq;\btheta_{\mathcal{I}})) \\
    &+ E_{\mathrm{elastic}}(\bx_t(\bz_t(\bq;\btheta_{\mathcal{I}})), \bX) \ .
\end{split}
\end{equation}
As demonstrated by the results, our physics-based loss function significantly improves the generalization ability of the trained network. For rod examples, we exclude the boundary condition term as it is enforced as hard constraints. For cloth and solid examples, we use a quadratic penalty potential, $E_{\mathrm{bc}}(\bx_t) = w_{bc}||\bx_t - \bx_t^*||^2_2$, to enforce boundary conditions, where $\bx_t^*$ is the target position at time step $t$.
\paragraph{Data Augmentation}
During training, we use ground truth latents from training data and predict latents only for one step. This by itself cannot guarantee stability during auto-regressive inference.
While one alternative is to predict multiple steps during training~\cite{grigorev2023hood}, the complexity of the training graph can increase significantly.
\par
We propose a simpler alternative without the need to predict longer time horizons during training.
Inspired by works on learning integrators in full space~\cite{sanchez2020learning,pfafflearning}, we augment our training data with the same amount of perturbed data.
Specifically, for each batch, we add random noise sampled from a uniform distribution to the previous two latents scaled by 10\% of the standard deviation of that batch while keeping the boundary conditions unchanged. Concretely, the integrator network also predicts $\tilde{\bz}_t$ given the perturbed inputs $\tilde{\bz}_{t-1}$ and $\tilde{\bz}_{t-2}$ 
\begin{equation}
    \tilde{\bz}_t = \mathcal{I}( \tilde{\bz}_{t-1}, \tilde{\bz}_{t-2}, \bp_t, \bp_{t-1},\bp_{t-2} ; \btheta_{\mathcal{I}}) \ .
\end{equation}
The same loss function (Eqn.~\ref{eqn:training}) is applied to $\tilde{\bz}_t$.
This training noise allows for efficient sampling of states that are not at dynamic equilibrium, and we are able to achieve this without additional data generation, thanks to our self-supervised setup.
This enables stable autoregressive inference for thousands of frames. 

\paragraph{Data Balancing}
We observe that different systems can have significantly varying energy levels, which impact their optimization. In particular, the higher energy state tends to dominate the gradient, causing errors in the lower energy systems. To illustrate this, consider how changes in boundary conditions affect a system's total energy. Small changes might only slightly increase the energy, but larger changes can cause significant jumps in the inertial and elastic potential. To mitigate this data imbalance, we employ a normalization strategy. Specifically, we normalize the energies within batches using the averaged velocity derived from the full space positions of the latents of the previous two steps. This approach helps stabilize the optimization process and ensures more accurate results in diverse energy states.

The normalized loss per data point $\Bar{\mathcal{L}}_i$ is then 
\begin{equation}
    \Bar{\mathcal{L}}^i = \frac{\mathcal{L}^i}{|\bv_{t-1}|} \ ,
\end{equation}
where $\bv_{t-1} = \frac{1}{N}\sum_j^N \frac{\bx^j_{t-1} - \bx^j_{t-2}}{\dt}$ and $N$ is the total number of vertices in full space.
\paragraph{Hyperparameters \& Implementation Details}
Due to the physical nature of our supervision, we do not need to balance our loss terms, which reduces the need for hyperparameter tuning. We use $10^{5}$ as our penalty weights $w_{bc}$ for enforcing Dirichlet boundary conditions for all examples. The parameters in the physics-based loss terms correspond to the real-world parameters of the simulated objects. For hair, we assume a density of $1.32 \text{g}/\text{cm}^3$, and a radius of $0.7 \text{mm}$. We use $4 \times 10^{7}$ GPa as Young's modulus and compute the stretching and bending parameters based on Bergou~\etal~\shortcite{bergou2010discrete}. For cloth, we use a Young's Modulus of $10^{6}$ GPa and a Poisson ratio of 0.45. We assume the cloth has a thickness of 0.3 cm and a density of $1.5 \times 10^{3}$ kg/m$^3$. Finally, for our deformable solids, we use a Young's Modulus of $10^{7}$ GPa, a Poisson ratio of 0.48, and assume unit mass density.    
\par
Our full space simulation framework is adapted from an open-source C++ simulation framework~\cite{wukong}. 
We implement the autoencoder and integrator training in PyTorch~\cite{imambi2021pytorch}.
Our autoencoder uses multilayer perceptron (MLP) based ResNet~\cite{he2016deep} with batch normalization and Swish activation functions. The encoder and decoder are symmetric in layers and number of neurons.
Our neural integrator also leverages an MLP-based network with Swish activation functions. The exact network architecture can be seen in Table~\ref{tab:cubature_comparison}.
For training both the autoencoder and the neural integrator we use a learning rate of $10^{-4}$. We train our autoencoders for 20,000 epochs using a batch size of 500. 
\section{Results} 
In this section, we demonstrate the versatility of our approach by showcasing its applicability to various domains, including discrete elastic rods, discrete elastic shells, and volumetric finite elements. Our method is capable of simulating highly nonlinear elastic deformations in real-time on a CPU with plausible dynamics. Furthermore, we emphasize the importance of two key components in our method: training noise and data balancing, both of which play crucial roles in ensuring stability and generating plausible simulation outcomes.
\subsection{Rod Simulation}
\begin{figure}[h]
    \centering
    \includegraphics[width=\linewidth]{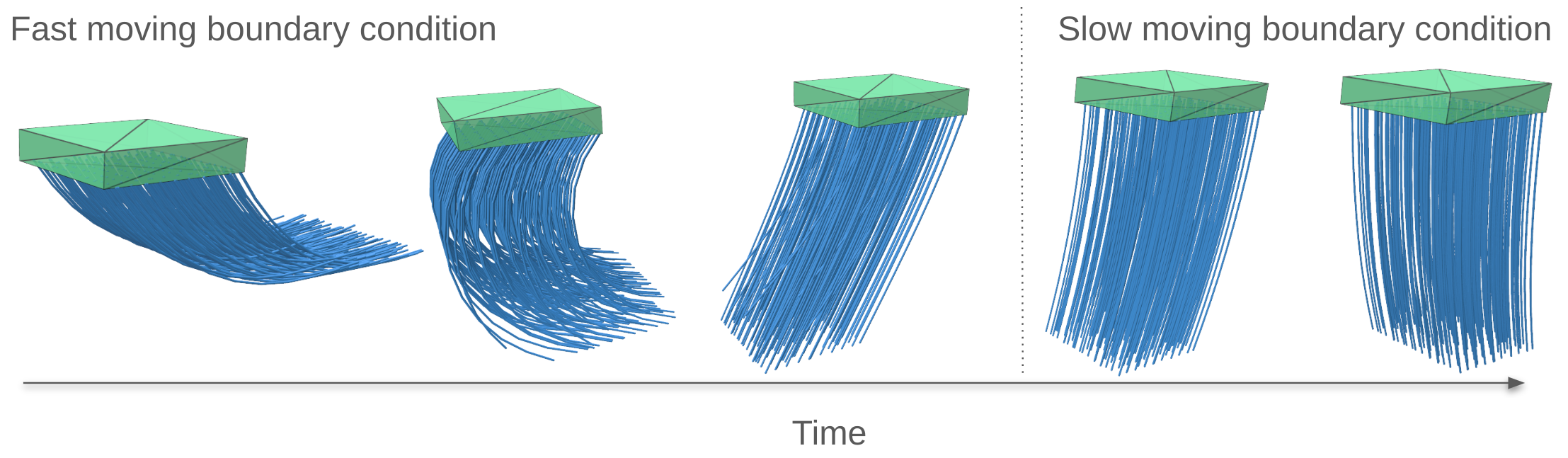}
    \caption{\textbf{Accelerating and Decelerating Rods:} Thanks to our self-supervised learning strategy, the rods can adapt to different velocities and accelerations unseen during training and produce plausible simulations.}
    \label{fig:rod_translation}
\end{figure}

In the first two sets of experiments, we show that our approach enables real-time simulation for discrete elastic rods. We begin by anchoring the rods at their root vertices and then applying a horizontal translation in the image plane (see Fig.~\ref{fig:rod_translation}). During training, we generate a sequence of examples with different but constant velocities. The rods are moved in opposite directions at 10-frame intervals for a total of 100 frames.  Our training data consists of 12 different sequences with boundary conditions moving at velocities ranging from 10 m/s to 120 m/s with increments of 10 m/s. At test time, we start with a fast and out-of-distribution velocity and suddenly decrease it by an order of magnitude. As can be seen from this sequence, our networks smoothly adapt to these changes in boundary conditions.
\par
\begin{figure}[h]
    \centering
    \includegraphics[width=\linewidth]{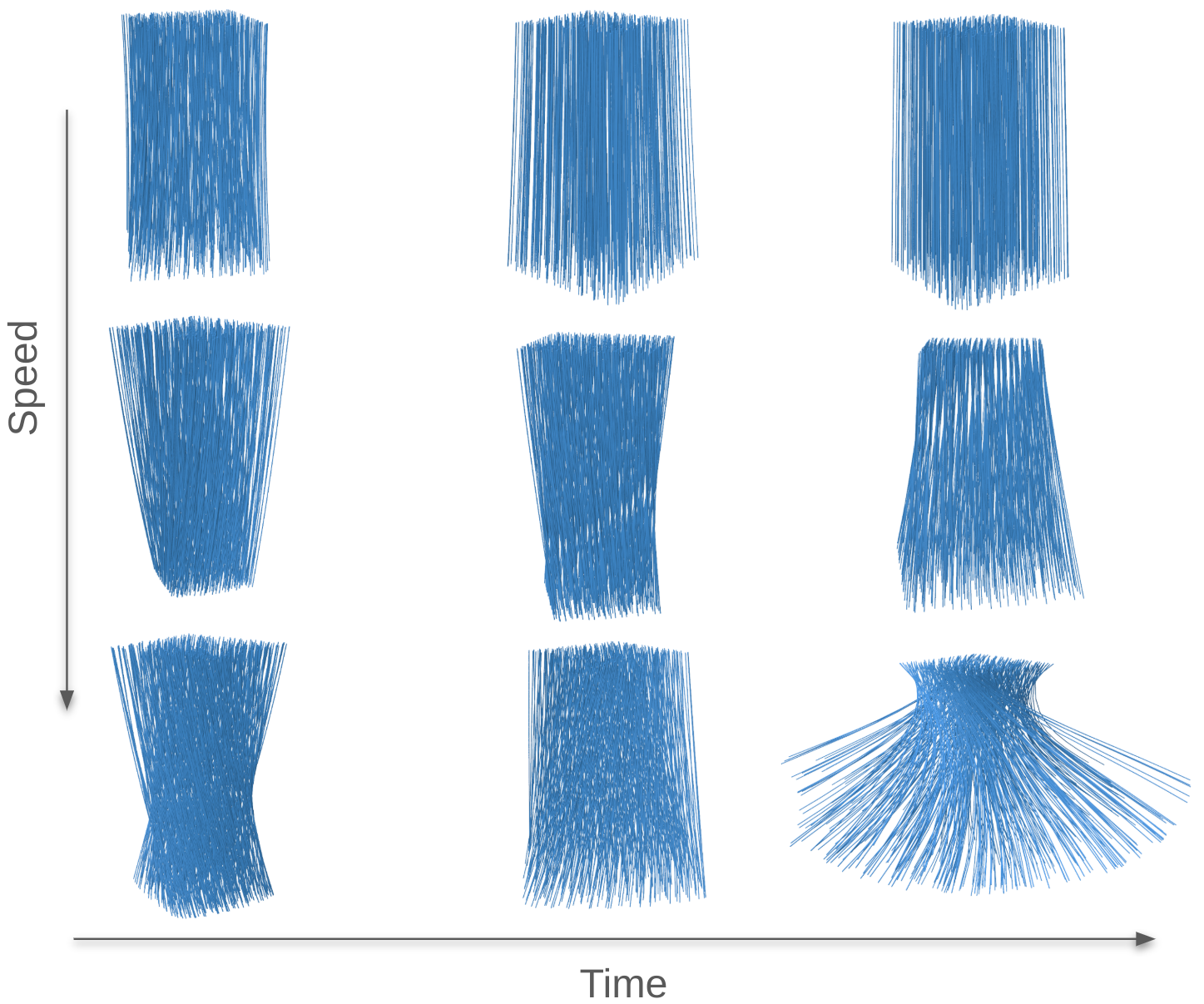}
    \caption{\textbf{Rods Under Rotation Boundary Conditions:} Our method produces plausible dynamics when adapting to changes in different boundary conditions, \ie rods swing much higher at large rotation speeds (third row) but do not accumulate large momentum when the rotation speed is small to medium (first and second row).}
    \label{fig:rod_rotation}
\end{figure}

In the second example, we rotate the same group of rods under different angular velocities. We create the training set by rotating the group of rods using different but constant velocities. We first test the resulting network using both interpolated velocities in between training data points (first and second rows) and outside of the training distribution (last row). Our training data consists of seven simulation sequences, each containing 100 frames. At each sequence, we vary the angular velocities from 0.1 $\degree$/s to 0.4 $\degree$/s in increments of 0.05 $\degree$/s. As can be seen from Fig.~\ref{fig:rod_rotation}, our method is able to adapt to these changes and produces plausible and stable dynamics.

\begin{figure}[h]
    \centering
    \includegraphics[width=\linewidth]{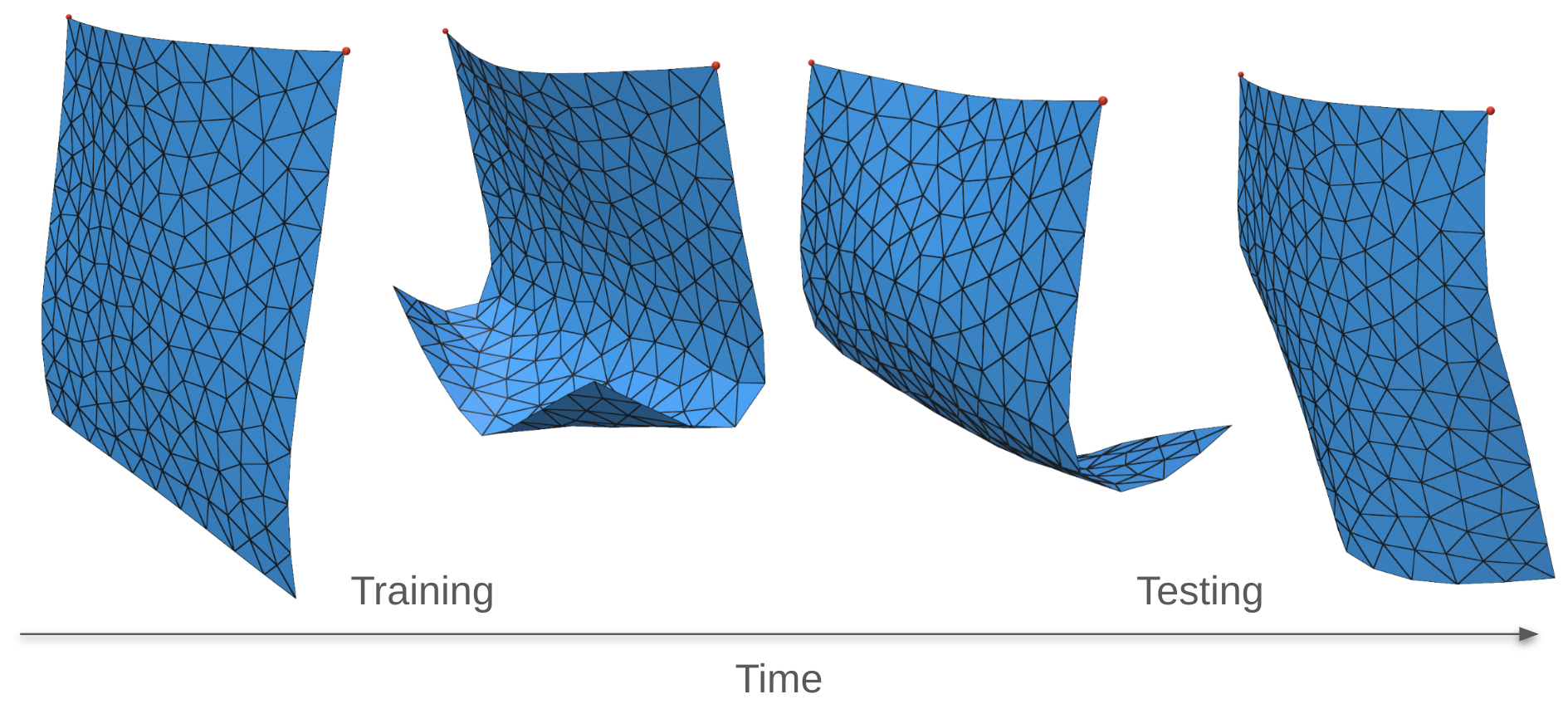}
    \caption{\textbf{Moving Cloth:} Two pinned corners of a piece of cloth are being moved along a linear trajectory. Our method is able to generate stable simulation outcomes for a longer time horizon, despite being trained on only half of the sequence.}
    \label{fig:discrete_elastic_shell}
\end{figure}

\subsection{Shell Simulation}
We now turn to cloth simulation by applying our approach to simulate a discrete elastic shell by incorporating the StVK stretching energy and hinge-based bending energy under the influence of gravity. To generate training and testing motions, we pin two corners of the shell and animate them with varying velocities along a linear trajectory. We train the neural integrator using the first half of the motion sequence (750 frames) and test it on the latter half, as depicted in Fig.~\ref{fig:discrete_elastic_shell}. The results demonstrate that our neural integrator successfully generates plausible motion that adheres to the boundary conditions and completes the trajectory smoothly.
\begin{figure}[h]
    \centering
    \includegraphics[width=\linewidth]{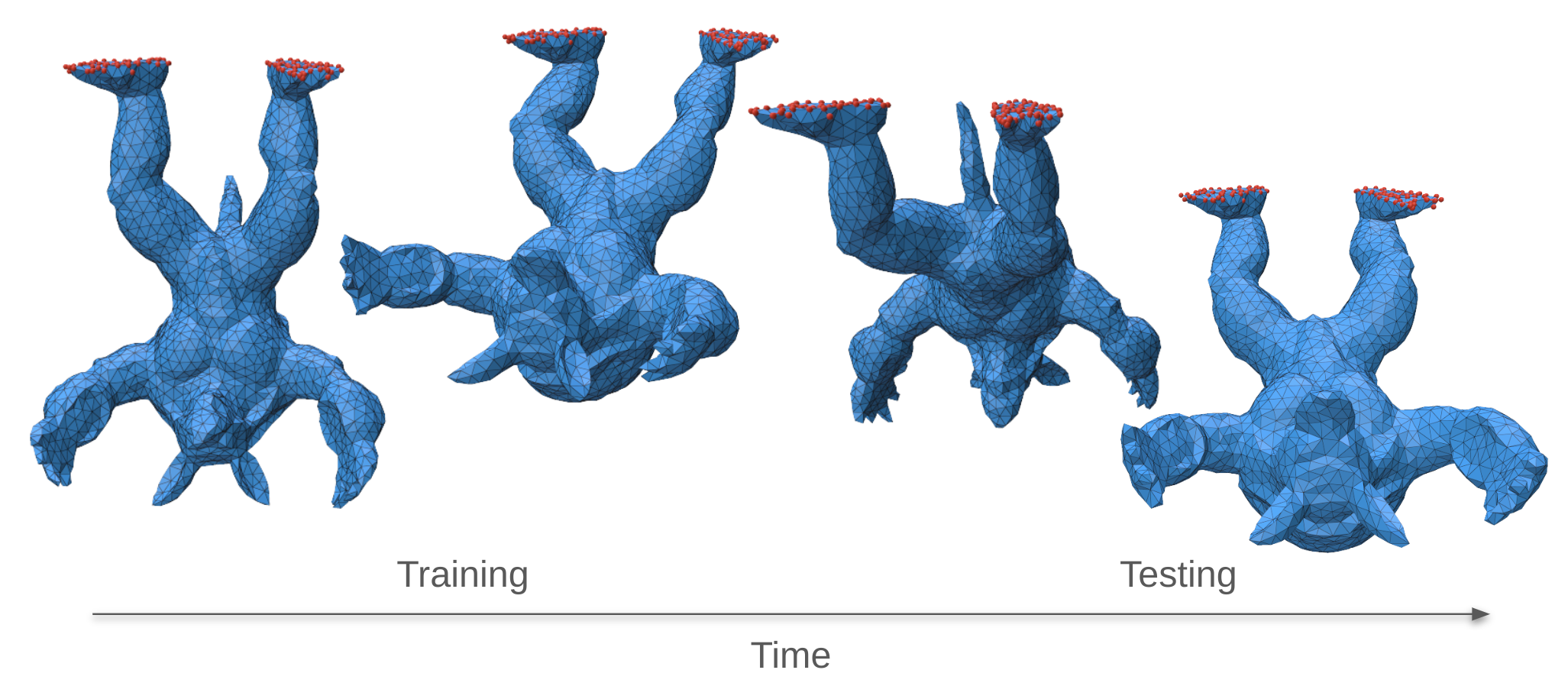}
    \caption{\textbf{Swinging Armadillo:} Our method simulates the elastic deformation of an upside-down armadillo under gravity, with the prescribed motion of the boundary vertices shown in red. The simulation accurately captures the nonlinear behavior of the dynamic motion and demonstrates generalization to unseen poses.}
    \label{fig:solid_sim}
\end{figure}

\subsection{Deformable Solid Simulation}
In the last sets of experiments (Fig.~\ref{fig:solid_sim}), we demonstrate results on 
deformable solids discretized by volumetric finite elements. We fix the upper points (as indicated by red spheres) of an upside-down armadillo and move it along a prescribed trajectory with varying velocities. 
We train on the initial 60\% of the sequence (780 frames) and test on the complete sequence (1300 frames). Our approach captures the nonlinear deformation in real time and remains stable even for unseen motions.

\begin{figure}[h]
    \centering
    \includegraphics[width=\linewidth]{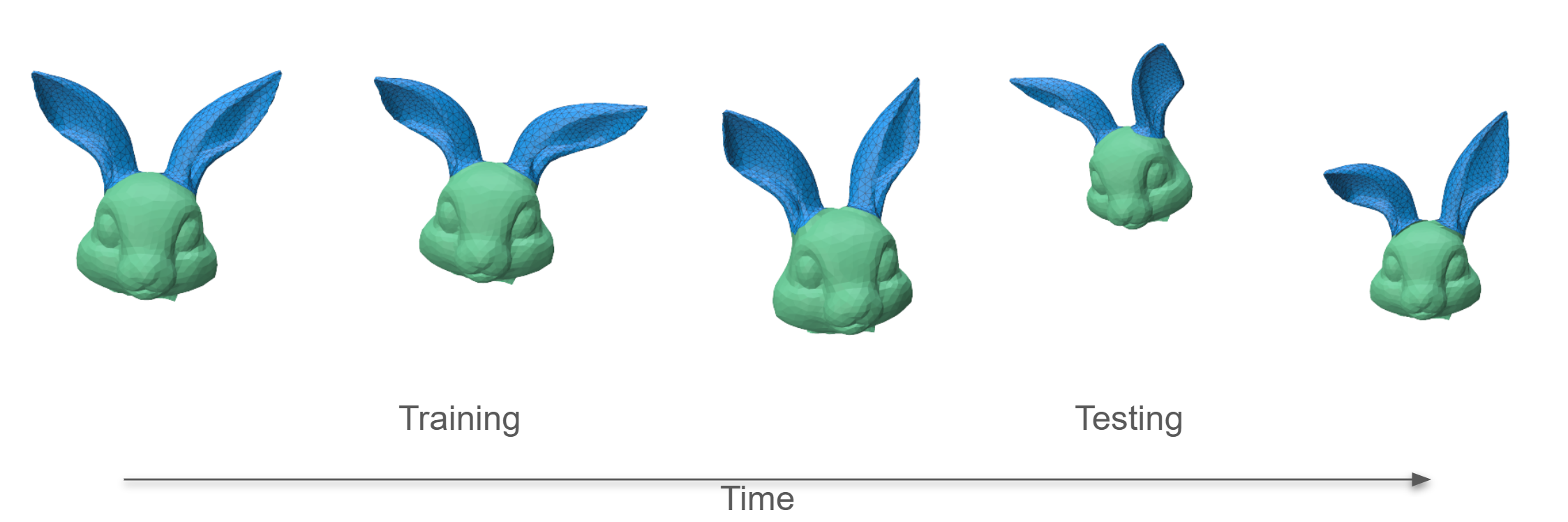}
    \caption{\textbf{Bunny Ears:} We simulate the elastic deformation of a pair of bunny ears attached to a rigid bunny head, with the prescribed motion of the head shown in green. The simulation captures the dynamic movement of the ears under gravity and predicts their motion on unseen sequences.}
    \label{fig:bunny_ears}
\end{figure}
We demonstrate another example in Fig.~\ref{fig:bunny_ears} by showcasing a pair of deformable bunny ears (blue) attached to a rigid bunny head (green) with a prescribed motion. We train on the first 60\% of the sequence, which consists of 540 frames, and test on the complete sequence, comprising 900 frames. Our method captures the dynamic motion of the ears under gravity and the change in motion of the bunny head, highlighting its effectiveness in handling complex scenarios involving both rigid and deformable objects.

\subsection{Comparisons}
\paragraph{Comparisons with Offline Simulation}
We compare our neural integrator with our reference offline simulation. As can be seen in Table~\ref{tab:comparison_offline_sim}, our method improves the performance by 3 orders of magnitude. All timings are reported on CPU. 
\begin{table}[h]
    \caption{\textbf{Quantitative comparison against reference offline simulation.} By improving the performance by orders of magnitudes, our method achieves real-time performance on CPU.}
    \centering
    \begin{tabular}{p{2.4cm} p{1.4cm} p{1.2cm} p{1.4cm} p{1.5cm} p{1.5cm} p{1.8cm}}
    \toprule
       Examples  &  Full Space DoF & Subspace DoF & Simulation & Ours Integration & Ours Decoder & \textbf{Ours Total}\\
       \midrule
       Hair Rotation  &  10,800 & 4 & 268 ms & 0.264 ms & 0.456 ms & \textbf{0.720 ms} \\
       Hair Translation  &  5,400 & 4 & 208 ms & 0.172 ms & 0.386 ms & \textbf{0.558 ms}\\
       \midrule
       Cloth   &  576 & 4 & 18.83 ms & 0.109 ms & 0.163 ms & \textbf{0.272 ms}\\
       \midrule
       
       Armadillo  &  10,020 & 12 & 136.8 ms & 0.253 ms & 0.236 ms & \textbf{0.489 ms}\\
       \midrule
       Bunny Ears & 8,046 & 8 & 112.6 ms & 0.228 ms & 0.198 ms & \textbf{0.426 ms} \\
       \bottomrule
       
    \end{tabular}
    \label{tab:comparison_offline_sim}
\end{table}

\paragraph{Comparisons with Latent Space Integrator}
While linear subspace simulation methods benefit from a constant map from reduced to full space, they require more bases for the same reconstruction quality~\cite{fulton2019latent}. We therefore favor nonlinear subspace simulation methods and compared our method with state-of-the-art methods by \citet{fulton2019latent} and \citet{shen2021high}. Both methods leverage autoencoders to construct nonlinear spaces and perform optimization-based integration in the subspace. At inference time, both approaches require full space reconstructions from latent codes. The energies as well as their derivatives for optimization are computed at selected and optimized \textit{cubature} points. On the contrary, our method entirely bypasses these computations by predicting next-step latent codes directly using a neural integrator. In this section, we show that our approach is faster to compute when compared to even a single computation of the energy and derivative required for a single \textit{cubature} point. In our implementation, we do not explicitly construct the Jacobian matrix, but instead rely on a convenient Jacobian-vector product function provided by PyTorch. This timing is reported in the second-to-last column of Table~\ref{tab:cubature_comparison}.
\begin{table*}[h]
    \caption{\textbf{Quantitative comparison against latent space integrator.} We compare our complete method against the simplest setting in an optimization-based latent space integration framework, \ie computing the Jacobian vector product for a single full space vertex \wrt the latent space variables once. In reality, more than one cubature is required, and computing this gradient more than once is necessary. Nonetheless, we show that our complete approach is already faster than this single operation. }
    \centering
    \begin{tabular}{p{2.4cm} p{1.4cm} p{1.2cm} p{1.4cm} p{1.8cm} p{1.5cm} p{1.4cm}}
    \toprule
       Examples  &  Full Space DoF & Subspace DoF & Decoder Layers & Integrator Layers & Computing Jvp & \textbf{Ours} \\
       \midrule
       Hair Rotation  &  10,800 & 4 & 512, 512, 512, 50 & 512, 512, 512 & 2.19 ms & \textbf{0.720 ms} \\
       Hair Translation  &  5,400 & 4 & 256, 256, 256, 50 & 256, 256, 256 & 1.51 ms & \textbf{0.558 ms} \\
       \midrule
       Cloth  &  576 & 4 & 200, 200, 200, 50 & 32, 32 & 0.564 ms & \textbf{0.272 ms} \\
       \midrule
       Armadillo  &  10,020 & 12 & 200, 200, 200, 50 & 256, 256, 256, 256, 256 & 0.649 ms & \textbf{0.489 ms} \\
       \midrule
       Bunny Ears & 8,046  & 8 & 200, 200, 200, 50 & 256, 256, 256, 256, 256 & 0.632 ms & \textbf{0.426 ms} \\
       \bottomrule
    \end{tabular}
    \label{tab:cubature_comparison}
\end{table*}

\begin{figure*}[h]
    \centering
    \includegraphics[width=\linewidth]{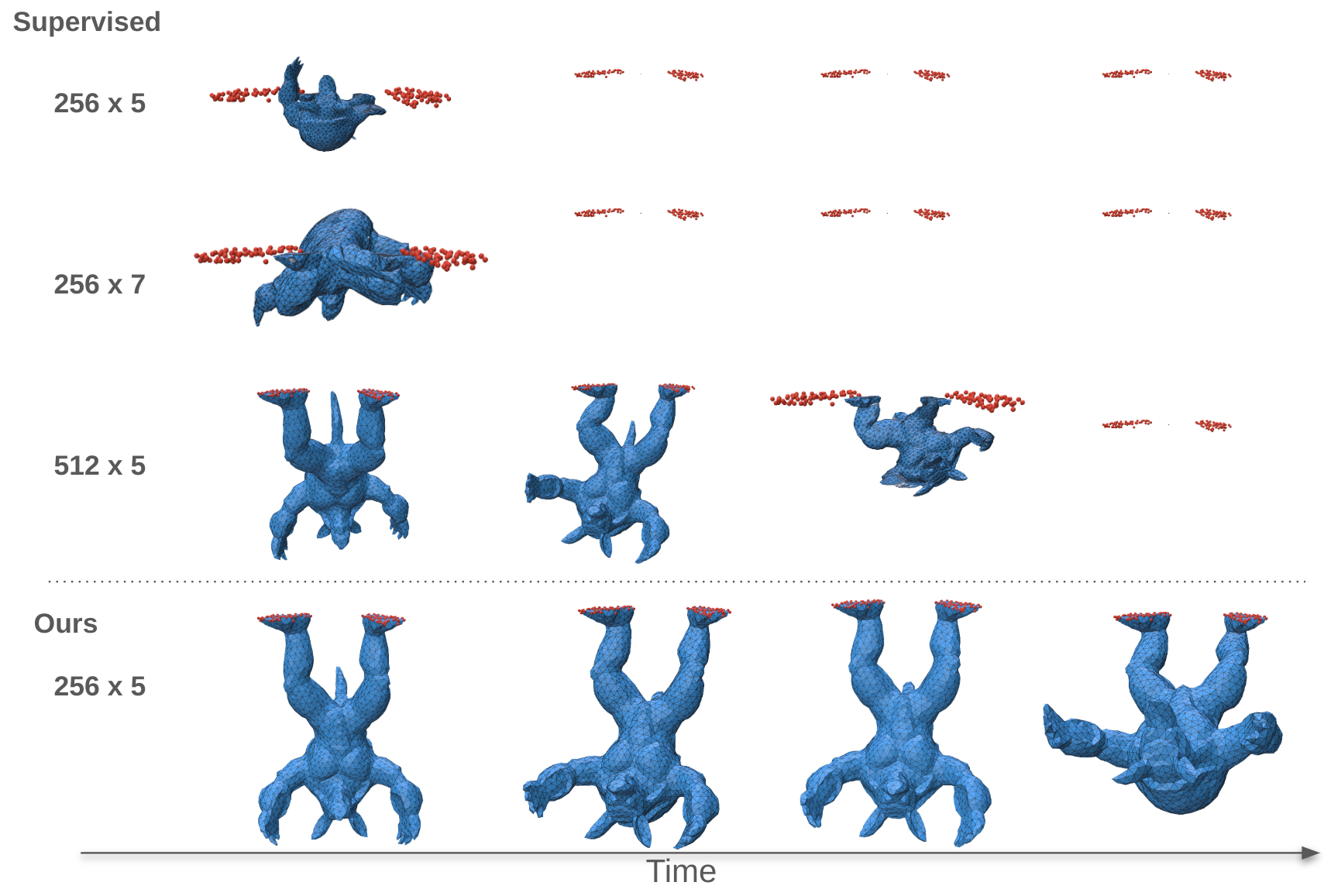}
    \caption{\textbf{Autoregressive Inference:} We compare our approach with a supervised learning strategy. Supervised learning lacks stability when it comes to long rollouts due to per-step error accumulation. While increasing network capacity mitigates this issue to some extent (\textit{row 1-3}), it diverges eventually for a thousand-frame sequence. On the contrary, our method (\textit{row 4}) enables stable simulation roll outs with plausible dynamics.}
    \label{fig:comparison_supervised}
\end{figure*}
\subsection{Ablation Studies}
\begin{figure}[h]
    \centering
    \includegraphics[width=\linewidth]{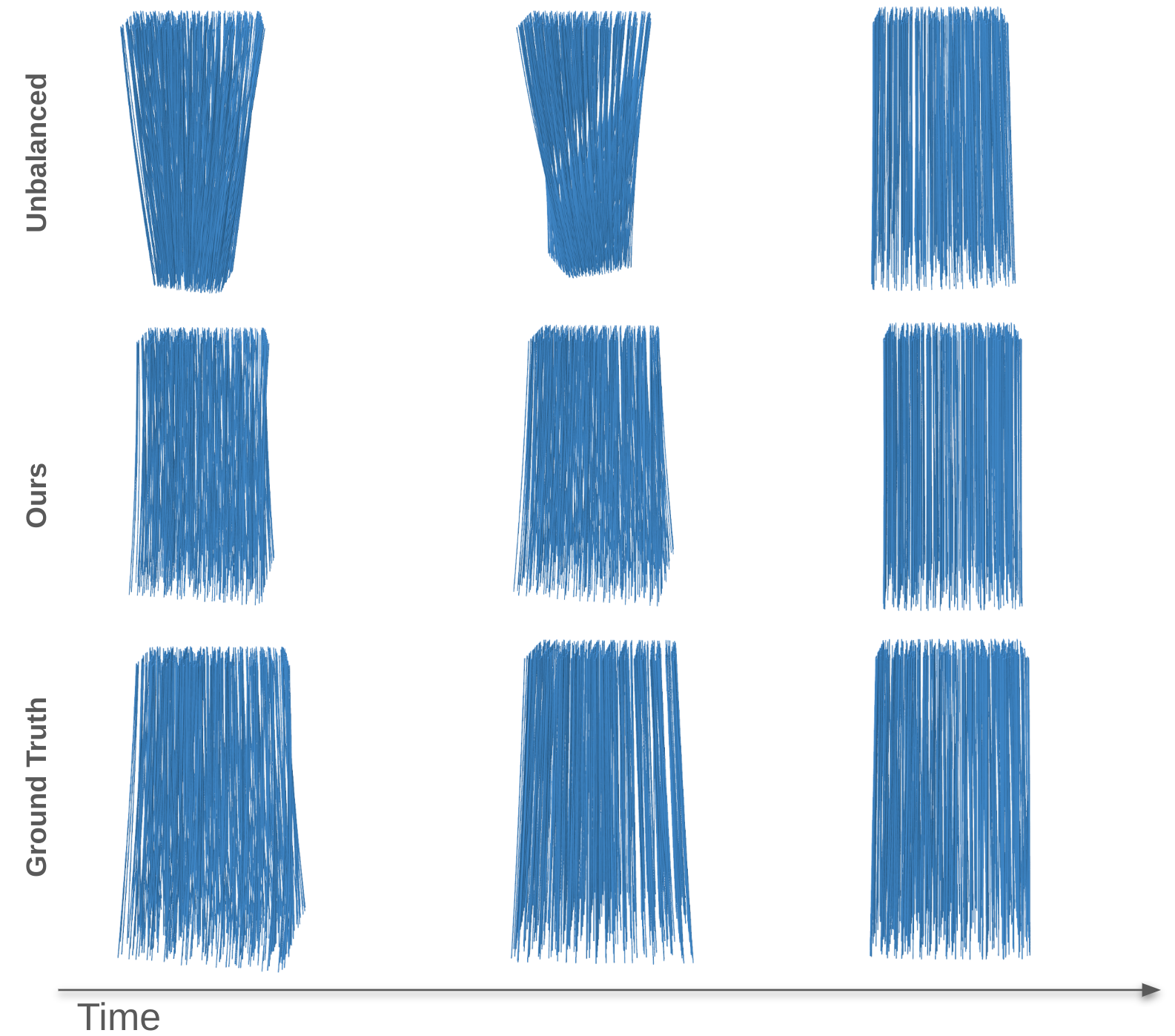}
    \caption{\textbf{Data Balancing:} Without balancing data between different system energies, the training process tends to favor high energy states due to the correspondingly large gradients. Using our balanced loss function improves the prediction, especially on low energy states. }
    \label{fig:data_balancing}
\end{figure}

\begin{figure}[h]
    \centering
    \includegraphics[width=0.9\linewidth]{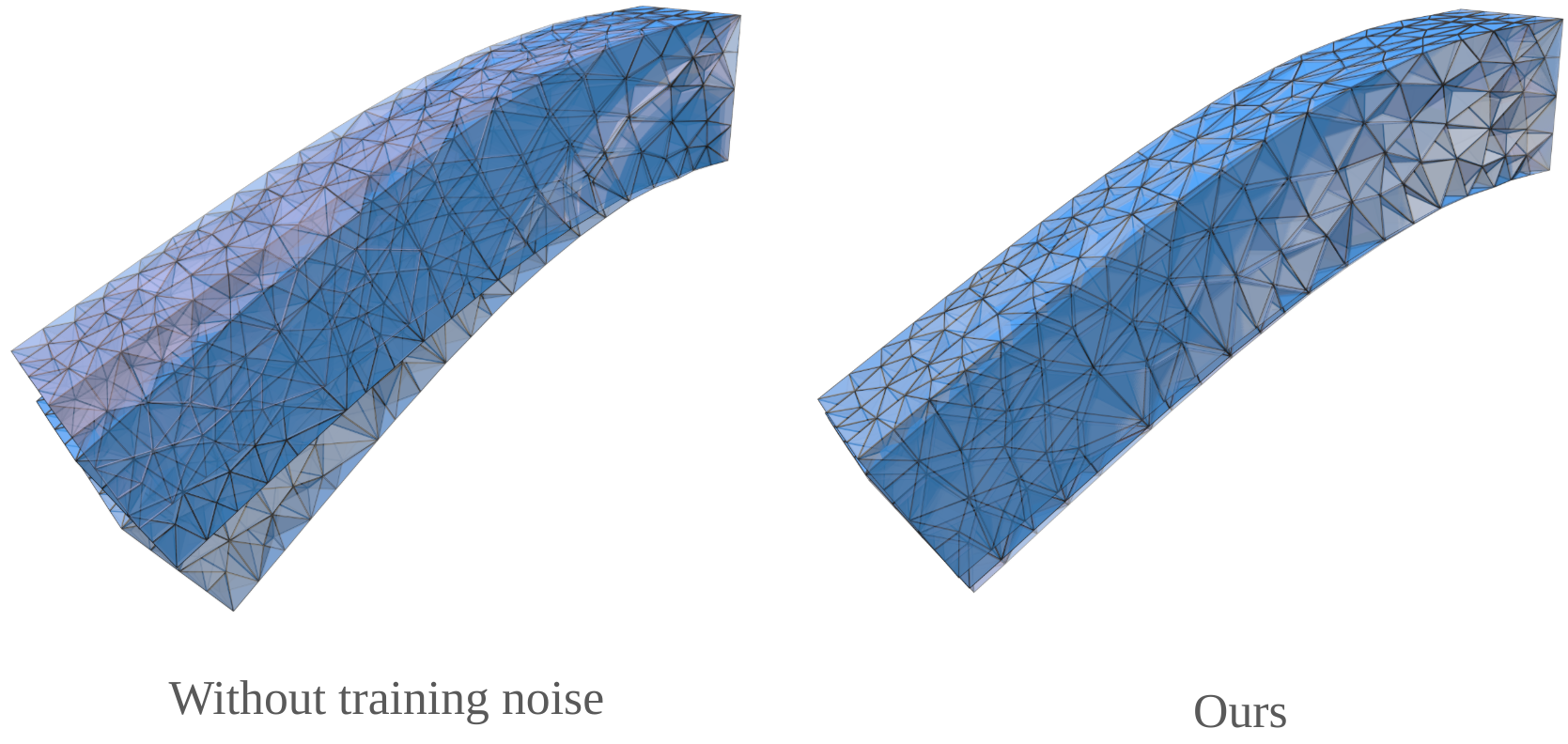}
    \caption{\textbf{Steady State Solution}: A cantilever beam clamped from one end is deformed under gravity. After training on the initial 100 frames of the simulation sequence, we continue the inference of the network for an additional 200 frames. The last frames of the inference are shown with decreasing transparency. Without training noise, the network prediction oscillates around the steady state solution (\textit{left}), whereas the training noise facilitates the network converging to the ground truth solution (\textit{right}).}
    \label{fig:ablation_noise_beam}
\end{figure}

\paragraph{Comparison with Supervised Learning}
We attribute robustness and stability to our self-supervised training strategy. In this example (Fig.\ref{fig:comparison_supervised}), we compare the rollout stability of our method with that of a supervised strategy. The first and last row use identical network parameters, while the only difference is the loss function. In the supervised setting, while the network converges to its local minimum, the $L_2$ error is still non-zero. During inference, only the first two frames precisely match the training data, and the per-time-step prediction error rapidly deviates from the training data distribution. Consequently, errors accumulate, causing autoregressive inference to diverge in fewer than 100 frames. While the $L_2$ error can be reduced by increased network capacity (\textit{row 2-3}), the supervised approach still suffers from stability issues. Our method is robust to rollout errors and remains stable after thousands of steps. For our method, we use an MLP of 5 layers of 256 neurons. For the supervised setting, we include the number of layers and the number of neurons per layer on the left. 
\paragraph{Data Balancing}
While our physics-based loss function automatically accounts for the weighting of different terms, data within each batch can still have energies in different value ranges, \eg, rapidly changing boundary conditions can lead to comparatively large momentum and elastic energies and vice versa. As can be seen in Fig.~\ref{fig:data_balancing}, without data balancing (\textit{row 1}), the rods curl much more inward compared to the ground truth data (\textit{row 3}) whereas our approach reproduces the desired motion (\textit{row 2}).
\paragraph{Training Noise}
In this example (Fig.~\ref{fig:ablation_noise_beam}), we show that adding noise during training helps prevent overfitting. We simulate a cantilever beam under gravity load with one end being fixed. With first-order implicit Euler integration, the simulation converges to a steady-state solution. We train our network using the first 100 frames of the sequence with and without training noise. During testing, we roll out 200 more frames than the training sequence. 
\par
In the second example, we study the effect of our training noise in a more complex setting. Our training data points only contain states that strictly satisfy the boundary conditions and obey the laws of physics. While our self-supervised loss function penalizes the deviations of equilibrium states for the third step, small errors in this prediction can still lead to divergence during auto-regressive inference (as can be seen from the first row). By adding noise to the input latents, we effectively sample out-of-balance states during training. Consequently, we achieve stable rollouts that satisfy the boundary conditions at all times (second row).
\begin{figure}[h]
    \centering
    \includegraphics[width=\linewidth]{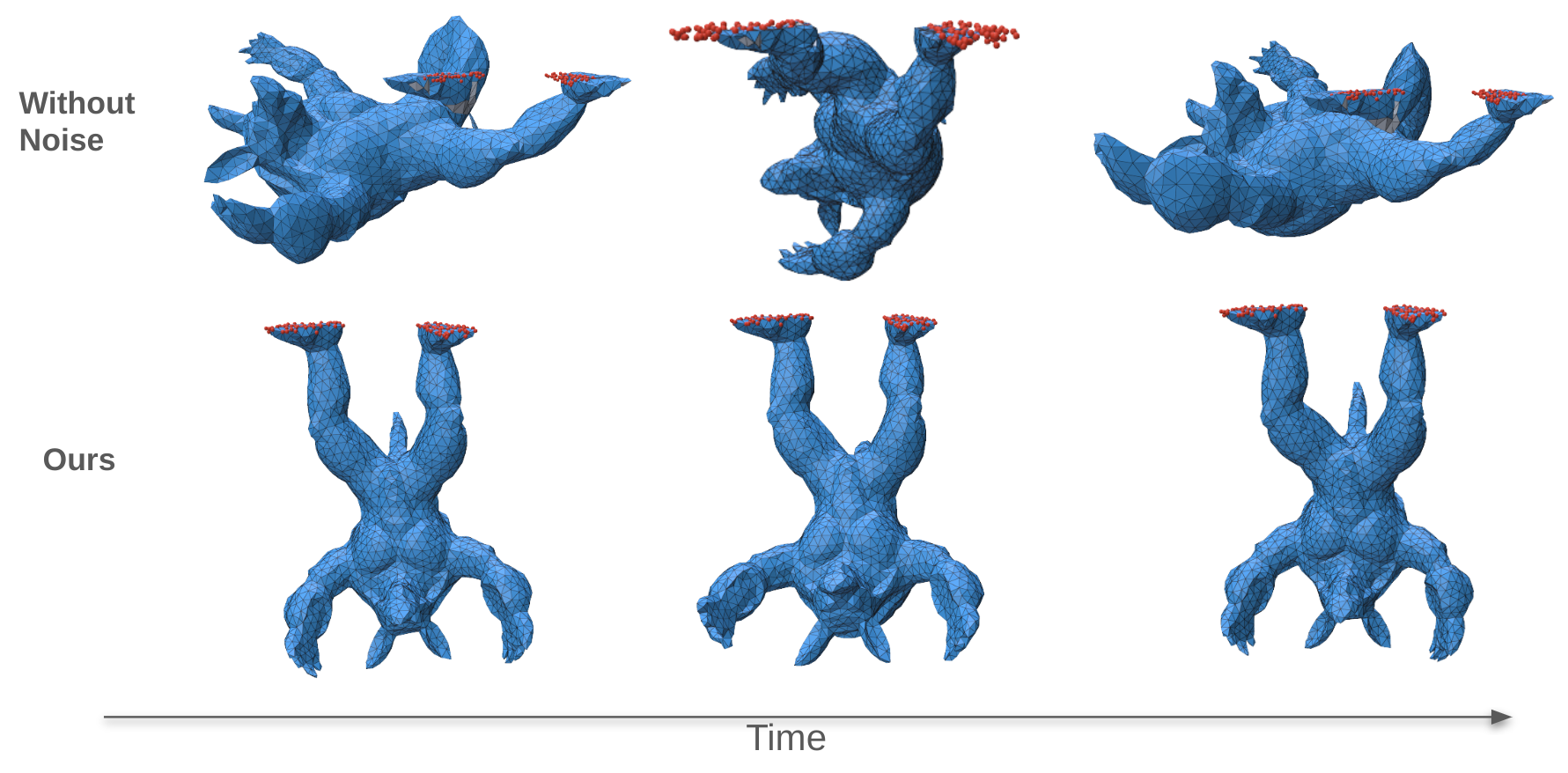}
    \caption{\textbf{Training Noise:} Without training noise, the network predictions gradually deviate from the boundary conditions and generate implausible states containing significantly large deformations. With training noise, our network predictions satisfy boundary conditions at all times.}
    \label{fig:ablation_noise_armadillo}
\end{figure}
\subsection{Autoencoder Training}
Our autoencoder reconstruction achieves a low error with no visible artifacts or visual errors. This high-fidelity reconstruction is crucial as larger errors can introduce spurious forces in full space, which hinders subsequent self-supervised training. The training time of the autoencoder ranges from 5 minutes to 30 minutes, depending on the complexity of the object geometry.
\subsection{Timings}
All timings in Tables~\ref{tab:comparison_offline_sim} and~\ref{tab:cubature_comparison} are reported from a workstation with an Nvidia 3080 GPU and an AMD Threadripper CPU. Our training process consists of 10,000 epochs, with each epoch taking less than 1 second to complete. The total training time is approximately 3 hours.
\section{Conclusion \& Future Work}
We have proposed a novel paradigm for real-time dynamic simulation of elastic objects in latent space. By leveraging a small neural network function exclusively in latent space, we accelerate offline simulation by orders of magnitude. We further proposed training noise and batch-weighted data normalization techniques to ensure stability for long-horizon autoregressive inference and data balancing. 
\par
\begin{figure}[h]
    \centering
    \includegraphics[width=\linewidth]{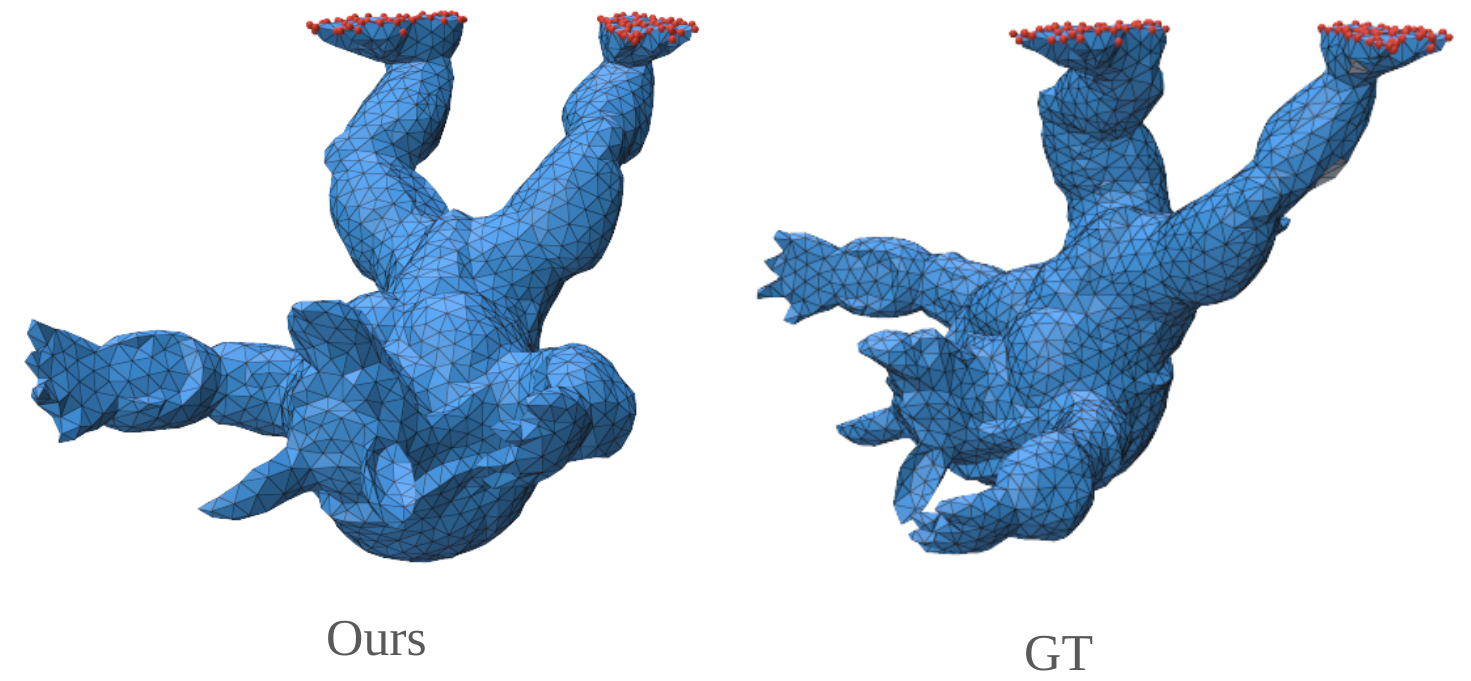}
    \caption{\textbf{Limitation:} our prediction (\textit{left}), albeit stable, exhibits kinetic energy loss compared to the ground truth data (\textit{right}).}
    \label{fig:lmitation}
\end{figure}

While our method enables real-time simulation at 30 FPS, it currently does not support varying frame rates at test time. 
Similar to existing subspace simulation methods, the nonlinear subspaces learned by our autoencoders are restricted to a single geometry and topology. Although beyond the scope of this work, applying topology independent model reduction techniques~\cite{modi2024simplicits} could be a promising direction for generalizing across geometries.
Our method could be extended to handle contact with a sufficiently smooth representation~\cite{du2024robust,larionov2021,li2020incremental}, and further reduction can be explored leveraging neural contact methods \cite{romero2021learning, romero2023learning, romero2022contact}.
Additionally, although material parameters could potentially be integrated as trainable variables, our approach does not allow online exploration of deformation behaviors across different materials. We have demonstrated that our method is more robust and stable compared to supervised alternatives; however, the kinetic energy in our simulated results remains lower than that observed in ground-truth simulation data (see Fig.~\ref{fig:lmitation}).
Our method currently works with simulation data only; extending the same paradigm to captured data, or reconstructed data given by point cloud or Gaussian splats, is another exciting direction for future exploration.

\begin{acks}
We thank Ryan Goldade for his assistance with the experiments and Wenzhao Xu for her help with the figures. We further thank the anonymous reviewers for their valuable feedback and suggestions.
\end{acks}
\bibliographystyle{ACM-Reference-Format}
\bibliography{reference}

\end{document}